\documentclass[onecolumn,showpacs,preprintnumbers,amsmath,amssymb]{revtex4}

\usepackage{graphicx}
\usepackage{dcolumn}
\usepackage{bm}
\usepackage{appendix}

\begin{document}

\preprint{APS/123-QED}

\title{Lindblad Plus From Feynman-Vernon}

\author{Jose A. Magpantay}
\email{jose.magpantay11@gmail.com}
\affiliation{Quezon City, Philippines}%

\date{\today}

\begin{abstract}
I show how the Lindblad Plus equation will follow from the Feynman-Vernon theory. The Plus refers to the inclusion of non-Markov processes in the Lindblad equation resulting in an integro-differential general master equation. The equivalence of this general master equation and not the Lindblad equation alone to the Feynman-Vernon theory should be expected because the sum over histories approach of the Feynman-Vernon theory clearly includes non-Markov processes, which Lindblad equation ignores.  This should close the seeming gap between these two approaches to quantum open systems.
\end{abstract}

\pacs{Valid PACS appear here}
\maketitle

\section{\label{sec:level1}Introduction}
There are two ways to deal with the quantum statistics of open systems, the first and derived earlier using path-integral techniques is the Feynman-Vernon (FV) theory \cite{Feynman} and the second derived more than a decade later is what became known as the Lindblad equation \cite{Lindblad}, which was also independently derived at about the same time by Gorini, Kossakowski and Sudarshan (GKS) \cite{Gorini}, from hereon will be referred to as the LGKS approach. The LGKS approach derives the replacement of the von Neuman equation for the density operator/matrix (see any statistical mechanics book, for example  \cite{Reichl}), under the general conditions of Hermiticity, Markov process and complete positivity. The Feynman-Vernon starts from the system-bath interaction and derive, using the path-integral method, the system's density matrix at any time t by integrating out the environment/bath degrees of freedom. In a specific case \cite{Caldeira}, the equation followed by the density matrix, which is a Fokker-Planck type of equation, can be derived from the path-integral result. However, still at this time there has been no paper that relates LGKS from FV. The link is crucial for it will provide how the terms that appear in the LGKS, the Hamiltonian and the Lindblad operator, are given in terms of the system-bath dynamics. But in a certain sense, the link does not seem to be warranted for the simple reason that FV, being a sum over histories approach would naturally include non-Markov effects while the LGKS is limited by the Markov process condition. 

In this paper, I will show how the FV theory is equivalent to what I call Lindblad Plus. This should lead to the identification of the terms that appear in LGKS, the Hamiltonian and the Lindbladian, and the memory terms in the Plus, in terms of the system/bath dynamic quantities - the system particle Lagrangian, the bath Lagrangian and the interaction between the two. This suggests that the two approaches to quantum statistics, the Feynman-Vernon and the Lindblad Plus are equivalent. To carry this out in a quite general way, I introduce the stationary phase method in the presence of a source. This will evaluate the path-integrals of the bath degrees of freedom subject to end point conditions. Integrating out the end point conditions will lead to an expression for the propagator for the density matrix, which can be compared with the path-integral of the Lindblad Plus. From these two expressions, the operative Hamiltonian, Lindbladian, and memory terms can be made. This is the main result in this work.

The paper is then arranged as follows. In Section II, I present the modification of the LGKS equation to include non-Markov processes, which I will call Lindblad Plus. Then I derive the path-integral that corresponds to this general master equation. In Section III, I summarize first the Feynman-Vernon theory. I point out that the propagator for the system density matrix is not Markovian. In Section IV, I make the connection between the Lindblad Plus and the FV and CL results using a particular example. Unfortunately, the example only has Plus terms, the Lindbladian is zero. In Section V, I show another way of evaluating the FV path-integrals using a stationary phase method with a source. This naturally leads to a transparent time-delay effects, which hints of memory effects. Integrating out the bath/environment initial and final degrees of freedom using the point mechanics example used by FV and CL yields the system density matrix, which can now be identified with the result of Section III. This  shows that the method of Section V can be used as a general approach to getting the propagator for the system's density matrix and then compare with the result of Section II to derive the Lindblad Plus terms - the operative Hamiltonian, the Lindbladian and the memory terms.  I conclude in Section VI what was achieved in this paper and options for future work.
\section{\label{sec:level2}Lindblad Plus}
The quantum statistical mechanics of a closed system begins from the von Neumann equation 
\begin{equation}\label{1}
i\hbar \dfrac{\partial \rho}{\partial t} = \left[ H, \rho \right] ,
\end{equation}
where $ \rho $ is the density operator and $ H $ is the Hamiltonian that describes the system's dynamics. The solution to equation (1) is given by
\begin{equation}\label{2}
\rho (t) = \exp {-\frac{i}{\hbar} H t } \rho (0) \exp {+\frac{i}{\hbar} H t },
\end{equation}
from which we get the time evolution of the density matrix $ \rho (x,y; t) $ as
\begin{equation}\label{3}
\begin{split}
\rho (x,y; t)& = \int dx'dy' K(x,x'; t) \rho (x',y'; 0) K^{*}(y',y; t)\\
                     & = \int dx'dy'J(x,y;x',y'; t) \rho(x',y'; 0).
\end{split}                      
\end{equation}
The $ K(x,x'; t) $ is the quantum propagator given by the path-integrals
\begin{equation}\label{4}
\begin{split}
K(x,x'; t)& = \int_{end pts} (d\tilde{x}) \exp {\frac{i}{\hbar} \int_{0}^{t} dt' L(\tilde{x},\dot{\tilde{x}})}\\
               & = \int_{end pts} (d\tilde{x})(dp_{x}) \exp {\frac{i}{\hbar} \int_{0}^{t} dt' \left[ p_{x}\dot{\tilde{x}} - H(\tilde{x},p_{x}) \right]},
\end{split}
\end{equation}
and $ J(x,y;x',y'; t) $ is the density matrix propagator and is given by
\begin{equation}\label{5}
\begin{split}
J(x,y;x',y'; t)& = K(x,x'; t) K^{*}(y',y; t)\\
                        & = \int_{end points} (d\tilde{x}) (d\tilde{y}) \exp {\frac{i}{\hbar} \int_{0}^{t} dt' \left[ L(\tilde{x},\dot{\tilde{x}}) - L(\tilde{y},\dot{\tilde{y}}) \right]} \\
                        & = \int_{end points} (d\tilde{x}) (d\tilde{y}) (dp_{x}) (dp_{y}) \exp {\frac{i}{\hbar} \int_{0}^{t} dt' \left( \left[ p_{x}\dot{\tilde{x}} - H(\tilde{x},p_{x}) \right] - \left[ p_{y}\dot{\tilde{y}} - H(\tilde{y},p_{y}) \right] \right)}.
\end{split}
\end{equation}
The first of the J expressions obviously show that J for closed systems is Markov, i.e., $ J(x,y; x',y'; t+ t') = \int d\bar{x} d\bar{y} J(x,y;\bar{x},\bar{y}; t,t+t') J(\bar{x},\bar{y};x',y'; 0, t) $. This is not satisfied by an open system as we will show later.

The end points in above refer to the coordinates end point conditions, i.e., $ \tilde{x}(t' = 0) = x', \tilde{x}(t' = t) = x $, similar end point conditions for $ \tilde{y}(t') $,  and no end point conditions on the momenta. 
For an open system, i.e., a system interacting with a bath/environment, there are two major approaches - the Feynman-Vernon theory and the Lindblad and Gorini - Kossakowski - Sudarshan theory (LGKS). The LGKS provides the analogous von Neumann type of equation for the system under conditions of (1) Markov process, (2) Hermiticity and (3) complete positivity and it is given by
\begin{equation}\label{6}
\begin{split}
i\hbar \dfrac{\partial \rho}{\partial t}& = [H,\rho] + i\textbf{L} \rho \textbf{L}^{\dagger} - \frac{i}{2}\left( \textbf{L}^{\dagger}\textbf{L}\rho + \rho \textbf{L}^{\dagger}\textbf{L} \right) \\
                                                                     & = H_{eff} \rho - \rho H^{\dagger}_{eff} + i \textit{L} \rho \textit{L}^{\dagger},
\end{split}
\end{equation}
where $ \textbf{L} $ is the Lindbladian and $ H_{eff} $ is given by
\begin{equation}\label{7}
H_{eff} = H - \frac{i}{2} \textbf{L}^{\dagger}\textbf{L}
\end{equation}
The second form of the LGKS equation is due to Struntz \cite{Struntz}, which provided the path integral for the time differential evolution operator of the system's density operator, which is the analogue of equations (3) and (4) for a closed system.  This form is significant for it will guide the inclusion of memory effects, the non-Markovian term, which must be present in general for a system interacting with a bath/environment for an extended period of time. This is what I will provide in the later part of this section. But first, the path-integral version of equation (6) is given in \cite{Struntz} as
\begin{equation}\label{8}
\rho(x,y; t) = \int dx'dy' J_{L}(x,y; x',y'; t) \rho(x',y'; 0), 
\end{equation}
where the four-point plus time function $ J_{L} $ is the propagator for the density matrix of a Lindblad system and it is given by the system's coordinates and momenta path-integrals
\begin{equation}\label{9}
\begin{split}
J_{L}(x,y; x',y'; t)&  = \int_{end points} (d\tilde{x}) (d\tilde{y}) (dp_{x}) (dp_{y})  \exp \frac{i}{\hbar} \int_{0}^{t} dt' \big\lbrace \left[ p_{x}\cdot\dot{\tilde{x}} - H(\tilde{x},p_{x}) \right] \\
 &\quad - \left[ p_{y}\cdot \dot{\tilde{y}} - H(\tilde{y},p_{y}) \right] -i \left[ (\textbf{L}^{\dagger}\textbf{L})(\tilde{x},p_{x}) + (\textbf{L}^{\dagger}\textbf{L})(\tilde{y},p_{y}) + \textbf{L}(\tilde{x},p_{x})\textbf{L}^{\dagger}(\tilde{y},p_{y}) \right]  \big\rbrace .
\end{split}
\end{equation}
The four-point plus time function $ J_{L} $  is the propagator for the density matrix and it is given by the system's coordinates and momenta path-integrals. To arrive at the Lindbladian terms in $ J_{L} $, I made use of equation (7). Note, the coordinates and momenta end points are the same as the end point conditions in equations (4) and (5). 

Before I include non-Markovian terms, I would like to point out that the propagator for the density matrix in LGKS is Markovian, as it should be because one of the conditions for the LGKS equation is the Markov condition. This follows from equation (9) where I expand $ J_{L}(x,y;x',y'; t+\tau) $ by (1) dividing the time integral at the RHS of (9) into  0 to t  then add the time integral from $ t to t+\tau $, (2) the path-integral measure  $ (dx') $ is divided into $ \left[ \prod dx'_{1}...dx'_{n}\right]  dx'_{n+1} \left[ \prod dx'_{n+2}...dx'_{N} \right] $,, and a similar expansion for $ (dy') $. There will be two path-integral measures, (1)  $ (dx') (dy') = \prod dx'_{1}...dx'_{n} dy'_{1}...dy'_{n} $ with end points $ x'(t' = 0) = x_{1}, x'(t' = t) = x'_{n+1} $ and a similar end point conditions for y', (2) $ (dx') (dy') = \prod dx'_{n+2}...dx'_{N} dy'_{n+2}...dy'_{N} $ with end points $ x'(t' = t) = x'_{n+1}, x'(t' = t+\tau) = x $ and a similar end point conditions for y'. Finally, the  integrations in $ dx'_{n+1} dy'_{n+1} $ will complete the Markovian condition $ J_{L}(x,y;x_{1},y_{1}; t+\tau) = \int dx'_{n+1} dy'_{n+1} J_{L}(x,y;x'_{n+1},y'_{n+1}; t+\tau, t) J_{L}(x'_{n+1},y'_{n+1}; x_{1},y_{1}; t, 0) $. 

It is important to point out that even though we have the path-integral for $ J_{L} $, we have not made a connection with the Feynman-Vernon theory because of two reasons. First, the LGKS equation is already assumed, implying knowledge of the operative H and the Lindbladian $ \textbf{L} $ (although there are a number of physical systems where these are known, see for example \cite{Chruscinski}). In general when a system interacts with a bath/environment, what is given are the separate system, bath dynamics and the interaction between the system's and bath degrees of freedom. The problem really is how to derive the operative Hamiltonian and the Lindbladian from these information. Second, the LGKS assumes the effective dynamics of the system from its interaction with the bath is Markov, which is hardly reasonable for a system that interacts with a bath/environment for an extended period of time. Given these arguments, it is not expected for the Feynman-Vernon theory to be related to equations (8) and (9). For this reason, I wanted to modify the LGKS equation to take into account the non-Markov processes that is expected to arise from a system-bath/environment interaction.

Consider the following integro-differential equation 
\begin{equation}\label{10}
i\hbar \dfrac{\partial \rho}{\partial t} =  H_{eff} \rho - \rho H^{\dagger}_{eff} + \textbf{L} \rho \textbf{L}^{\dagger} + \int_{0}^{t} d\tau \left[ M(t,\tau) \rho(\tau) - \rho(\tau)M^{\dagger}(t,\tau) + N(t,\tau)\rho(\tau)N^{\delta}(t,\tau) \right] ,
\end{equation}
The structure of the non-Markov term is the same as Lindblad's except for the non-locality in time. Also, note the $ N^{\delta} $ and not  the Hermitean adjoint $ N^{\dagger} $. This will become clear in a particular example later. The objective now is to determine the analogue of equation (9). The derivation starts from
\begin{equation}\label{11}
\begin{split}
\left\langle x|\rho(t+\vartriangle t) |y\right\rangle & = \int dx_{1}dy_{1}dp_{1}dp'_{1} \exp {\frac{i}{\hbar}\left[ p_{1}(x-x_{1}) - p'_{1}(y-y_{1}) \right]} \Big\lbrace  \Big[  1 - \frac{i}{\hbar}\vartriangle t \left(  H(\frac{x+x_{1}}{2},p_{1}) - H(\frac{y+y_{1}}{2},p'_{1}) \right)\\
& \quad  - \vartriangle t \frac{1}{\hbar} \left(-(\textbf{L}^{\dagger}\textbf{L})(\frac{x+x_{1}}{2},p_{1}) - (\textbf{L}^{\dagger}\textbf{L})(\frac{y+y_{1}}{2},p'_{1}) + \textbf{L}^{\dagger}(\frac{x+x_{1}}{2},p_{1})\textbf{L}(\frac{y+y_{1}}{2},p'_{1}) \right) \Big]  \left\langle x_{1}| \rho(t) |y_{1} \right\rangle \\ 
& \quad + \vartriangle t \frac{1}{\hbar} \int_{0}^{t} d\tau \Big[ M(\frac{x+x_{1}}{2},p_{1}; t,\tau) - M^{\dagger}(\frac{y+y_{1}}{2},p'_{1}; t,\tau) \\
& \quad + N(\frac{x+x_{1}}{2},p_{1}; t,\tau)N^{\delta}(\frac{y+y_{1}}{2},p'_{1}; t,\tau) \Big] \left\langle x_{1}| \rho(\tau) |y_{1}\right\rangle \Big\rbrace .
\end{split}
\end{equation}   
To derive the corresponding path integral expression for the general master equation given by equation (10), I follow the same steps that led to equations (8) and (9), i.e., divide the time interval $ [0,t] $ into N steps, from $ [0, \vartriangle t ], [ \vartriangle t, 2\vartriangle t ], ..., [(N-1)\vartriangle t, N\vartriangle t = t ] $. There are two important things to take note of - (1)  the $ n^{th} $ step density function step will require the $ (n-1)^{st} $ density function in the Markov part and the density function for all the previous steps in the non-Markovian terms, and (2) divide the time integral in the non-Markov part by steps in $ \vartriangle t $ and use the integration rule , i.e.,
\begin{subequations}\label{12}
\begin{gather}
\int_{0}^{n\vartriangle t} f(t') dt' = \int_{0}^{\vartriangle t} f(t') dt' + \int_{\vartriangle t}^{2\vartriangle t} f(t') dt' + ... \int_{(n-1)\vartriangle t}^{n\vartriangle t } f(t') dt', \label{first}\\
\int_{t_{1}}^{t_{2}} f(t') dt' = f(t_{2})(t_{2} - t_{1}),
\end{gather}
\end{subequations}
where equation (12b) is valid for two points infinitesimally separated. Using equations (12 a,b) , we find that the propagator for the density matrix is now given by (note, equation (8) is still valid, only equation (9) will change)
\begin{equation}\label{13}
\begin{split}
J_{LP}(x,y; x',y'; t)&  = \int_{end points} (d\tilde{x}) (d\tilde{y}) (dp_{x}) (dp_{y})  \exp \frac{i}{\hbar} \int_{0}^{t} dt' \Big\lbrace \left[ p_{x}\cdot\dot{\tilde{x}} - H(\tilde{x},p_{x}) \right] \\
 &\quad - \left[ p_{y}\cdot \dot{\tilde{y}} - H(\tilde{y},p_{y}) \right] -i \left[ (\textbf{L}^{\dagger}\textbf{L})(\tilde{x},p_{x}) + (\textbf{L}^{\dagger}\textbf{L})(\tilde{y},p_{y}) + \textbf{L}(\tilde{x},p_{x})\textbf{L}^{\dagger}(\tilde{y},p_{y}) \right] \\
 & \quad + \int_{0}^{t'}  dt'' \Big[ M(\tilde{x},p_{x}; t',t'') - M^{\dagger}(\tilde{y},p_{y}; t',t'') + N(\tilde{x},p_{x}; t',t'') N^{\delta}(\tilde{y},p_{y}; t',t'') \Big\rbrace ,
 \end{split}
 \end{equation}
 where the subscript LP in the above J stands for Lindblad Plus. The end points condition in equation (13) is the same as that in equation (9) and equation (5). The fact that the time integration in t'' is only from 0 to t'  follows from the rules given in equation (12). Also notice that the above J is clearly non-Markov. 

I will make use of equation (13) in identifying the operative Hamiltonian, the Lindbladian and the non-Markov terms when the propagator for the density matrix is derived starting from the system-bath/environment dynamics following the Feynman-Vernon procedure.
\section{\label{sec:level3}Feynman-Vernon Theory}
Consider a system, with coordinates $ x_{a}, a = 1,..,n $ and internal dynamics defined by the Lagrangian $ L_{s}(x_{a},\dot{x_{a}}) $, interacting with a bath/environment, assumed to be much larger than the system and at temperature T, with coordinates $ X_{i}, i = 1,..., N $ and $ N >> n $. The bath's dynamics is defined by $ L_{b}(X_{i},\dot{X_{i}}) $ and the system - bath interaction is defined by $ L_{int}(x_{a},X_{i}) $. It will be assumed that the interaction was turned on at $ t = 0 $. The combined system plus bath is a closed system and thus equations (1) to (4) are valid. The density matrix for the system plus bath is then given by
\begin{equation}\label{14}
\rho(x,X; y,Y; t) = \int dx'dX'dy'dY' K(x,X;x',X'; t) \rho(x',X'; y',Y'; 0) K^{*}(y',Y';y,Y; t),
\end{equation}
where the indices a and i are suppressed in the above equation for simplicity. Since the system, bath interaction is turned on only at $ t = 0 $, the initial density matrix factorizes  as
\begin{equation}\label{15}
\rho (x',X'; y',Y'; 0) = \rho_{s}(x',y'; 0) \rho_{b}(X',Y'; 0)
\end{equation}
where the subscripts s and b that goes with $ \rho $ at the right hand side of the above equation refer to system and bath respectively. 

Focusing on the system's density matrix, it is given by tracing over the bath's initial and final end points as given by
\begin{equation}\label{16}
\rho_{s}(x,y; t) = \int dX \rho(x,X;y,X; t).
\end{equation}
Using equations (14) and (15) in above, the result is
\begin{subequations}\label{17}
\begin{gather}
\rho_{s}(x,y; t) = \int dx' dy' J(x,y;x',y'; t)_{FV} \rho_{s}(x',y'; 0), \label{first}\\
J_{FV}(x,y;x',y'; t) = \int dX dX' dY' K(x,X;x',X'; t) \rho_{b}(X',Y'; 0) K^{*}(y',Y';y,X; t),
\end{gather}
\end{subequations}
where the subscript FV in above J stands for Feynman-Vernon and integrations in the above J are over the bath end point conditions (note, the end point condition at t is only X because of the tracing made). Before expressing this $ J_{FV} $ in terms of a path-integral, it is important to notice that it is not Markov because of the presence of the bath's $ \rho_{b}(X_{1},Y_{1}; 0) $ between the two K's. This should be expected because the bath particles incessant interaction with the system's degrees of freedom will imprint long term memory effects on the system. It is encouraging that the $ J_{FV} $ in equation (17b) has the same character, non-Markov, as the $ J_{LP} $ in the Lindblad Plus as given by equation (13).

The quantum propagator K for the system plus bath is given by
\begin{equation}\label{18}
\begin{split}
K(x,X;x',X'; t)& = \left\langle x X \Big\vert \exp {-\frac{i}{\hbar} H_{s+b} t} \Big\vert x'X' \right\rangle,\\
                         & = \int_{end points} (d\tilde{x}) (d\tilde{X}) \exp {\frac{i}{\hbar} \int_{0}^{t} dt' \left[ L_{s}(\tilde{x}) + L_{b}(\tilde{X}) + L_{int}(\tilde{x},\tilde{X}) \right] }.
\end{split}
\end{equation} 
The end points condition in above are $ \tilde{x}(t' = 0) = x', \tilde{x}(t' = t) = x, \tilde{X}(t' = 0) = X', \tilde{X}(t' = t) = X $, and there are similar conditions for $ \tilde{Y} $. Substituting equation (18) and a similar equation for $ K^{*} $ in equation (17b), the propagator for the distribution function of the system becomes
\begin{equation}\label{19}
J_{FV}(x,y;x',y'; t) = \int_{end points} (d\tilde{x}) (d\tilde{y}) \textbf{F}(\tilde{x},\tilde{y}) \exp {\frac{i}{\hbar}\int dt' \left[ L_{s}(\tilde{x}(t')) - L_{s}(\tilde{y}(t')) \right] }, 
\end{equation}
where the end points condition in above are $ \tilde{x}(t' = 0) = x', \tilde{x}(t' = t) = x $ and similar end point conditions for $ \tilde{y} $. The functional $ \textbf{F} $ is known as the influence functional and is given by the path-integral and ordinary integrals
\begin{equation}\label{20}
\begin{split}
\textbf{F}(\tilde{x},\tilde{y})& = 	\int dX dX' dY' \rho_{b}(X',Y'; 0) \int_{end points} (d\tilde{X}) (d\tilde{Y}) \exp \frac{i}{\hbar} \int dt' \Big\lbrace L_{b}(\tilde{X}(t')) + L_{int}(\tilde{x}(t'),\tilde{X}(t'))\\
&\quad - L_{b}(\tilde{y}(t')) -L_{int}(\tilde{y}(t'),\tilde{Y}(t')) \Big\rbrace 
\end{split}
\end{equation}
The end points in the above path-integral are defined by $ \tilde{X}(t' = 0) = X', \tilde{X}(t' = t) = X = \tilde{Y}(t' = t), \tilde{Y}(t' = 0) = Y' $, and the ordinary integrals are over the initial and final end points.

Equations (14) to (20) define the Feynman-Vernon theory. Given a specific system-bath dynamics, evaluating equations (20) and (19) and comparing with the Lindblad Plus path integral given by equation (13) will lead to the identification of the operative Hamiltonian, the Lindbladian and the non-Markov terms M and N. This is what will be done in the next section.
\section{\label{sec:level4}A Simple Example}
In this section I will relate the Lindblad Plus to the Feynman-Vernon theory through a particular example used by Feynman-Vernon and Caldeira-Legget \cite{Caldeira}. The system has a single particle with coordinate x and the bath is described by coordinates $ X_{i} $ with  $ i = 1,.., N $ with N taken to be a large number. The bath is assumed to be a system of independent harmonic oscillators each with a particular frequency $ \omega_{i} $ but all having the same mass m and the interaction of the particle with the bath is given by a simple product term. The Lagrangians are
\begin{subequations}\label{21} 
\begin{gather}
L_{b} = \frac{m}{2} \sum_{i = 1}^{N} \left[ \dot{X}_{i}^{2} - \omega_{i}^{2} X_{i}^{2} \right] , \label{first}\\
L_{int} = -x\sum_{i = 1}^{N} c_{i}X_{i},
\end{gather}
\end{subequations}
where $ c_{i} $ are constants.

I first isolate the path integral $ (d\tilde{X}) $ in equation (20), essentially defining how the system coordinate $ \tilde{x} $ dynamics is affected by the sum over the bath's degrees subject to the bath's end point conditions through the expression
\begin{equation}\label{22}
\kappa(\tilde{x}; X', X; t) = \int_{end points} (d\tilde{X}) \exp {\frac{i}{\hbar}} \int dt' \left[ L_{b}(\tilde{X}(t')) + L_{int}(\tilde{x}(t'),\tilde{X}(t'))\right].
\end{equation}
This had been carried out in equation (4.7b) of \cite{Feynman} and the result is
\begin{equation}\label{23}
\begin{split}
\kappa(\tilde{x}; X', X; t)& =  \exp \frac{i}{\hbar} \sum_{i = 0}^{N} \Big\lbrace \frac{m\omega_{i}}{2} (\cot \omega_{i}t) (X_{i}^{2} + X'_{i}{2}) - \dfrac{m\omega_{i}}{\sin \omega_{i}t} X_{i}X'_{i}\\ 
& \quad +c_{i} \int_{0}^{t} dt'  \left[- \dfrac{1}{\sin \omega_{i}t} X_{i} \sin \omega_{i}t' + X'_{i} \left( \cot \omega_{i}t \sin \omega_{i}t' - \cos \omega_{i}t' \right) \right] \tilde{x}(t')\\
& \quad +c_{i}^{2}  \frac{1}{m \omega_{i}} \dfrac{1}{\sin \omega_{i}t} \int_{0}^{t} dt' \int_{0}^{t'} dt'' \tilde{x}(t') \sin \omega_{i}(t - t') \sin \omega_{i}t'' \tilde{x}(t'') \Big\rbrace .
\end{split}
\end{equation}
There is a similar expression for the path-integral for $ (d\tilde{Y}) $. The last component of the influence functional $ \textbf{F} $ in equation (20) is the initial bath density matrix $ \rho_{b}(X'_{i},Y'_{i}; 0) $ and it is given in \cite{Caldeira}
\begin{equation}\label{24}
\rho_{b}(X',Y'; 0) = \prod_{i} \dfrac{m\omega_{i}}{2\pi \hbar \sinh \frac{\hbar\omega_{i}}{kT}} \exp {\left\lbrace -\dfrac{m\omega_{i}}{2\hbar \sinh \frac{\hbar\omega_{i}}{kT}} \left[ (X_{i}^{\prime 2} + Y_{i}^{\prime 2}) \cosh \frac{\hbar \omega_{i}}{kT} - 2X'_{i}Y'_{i} \right] \right\rbrace }
\end{equation}
The ordinary integrations over the bath end points $ X, X', Y' $ are simple. The details are given in Appendix A and upon substituting the above influence functional in equation (19) gives the propagator for the density matrix 
\begin{equation}\label{25}
\begin{split}
J(x,y;x',y'; t)& = \int_{end points} (d\tilde{x})(d\tilde{y}) \exp  \frac{i}{\hbar} \big\lbrace  \int_{0}^{t} dt' \left[ L_{s}(\tilde{x},\dot{\tilde{x}}) - L_{s}(\tilde{y},\dot{\tilde{y}}) \right] \\ 
& \quad + \int_{0}^{t} dt' \int_{0}^{t'} dt'' [ \tilde{x}(t') - \tilde{y}(t') ] \left( i \sum_{i} \dfrac{c_{i}^{2}}{m\omega_{i}} \left( \coth \frac{\hbar \omega_{i}}{k T} + \dfrac{1}{\sinh \frac{\hbar \omega_{i}}{kT} } \right) \cos \omega_{i}(t' - t'') \right) [ \tilde{x}(t'') - \tilde{y}(t'') ] \\
& \quad + \int_{0}^{t} dt' \int_{0}^{t'} dt'' [ \tilde{x}(t') - \tilde{y}(t') ] \left( \sum_{i} \frac{ c_{i}^{2}}{2m\omega_{i}} \sin \omega_{i}(t' - t'') \right) [ \tilde{x}(t'') + \tilde{y}(t'') ] \big\rbrace 
\end{split}
\end{equation}

To be able to compare this with equation (13) of Lindblad Plus, there are two things that must be done. First is to simplify equation (13) by assuming that the Lindbladian $ \textbf{L}(\tilde{x},p_{x}) $ and the non-Markov terms $ M(\tilde{x},p_{x}; t',t'') $ and $ N(\tilde{x},p_{x}; t',t'') $ are momentum independent. Integrating out the momenta, equation (13) simplifies to
\begin{equation}\label{26}
\begin{split}
J(x,y;x',y'; t)& =  \int_{end points} (d\tilde{x}) (d\tilde{y}) \exp \frac{i}{\hbar} \int_{0}^{t} dt' \Big\lbrace \left[ L'(\tilde{x},\dot{\tilde{x}}) - L'(\tilde{y},\dot{\tilde{y}}) \right]  -i \left[ (\textbf{L}^{\dagger}\textbf{L})(\tilde{x}) + (\textbf{L}^{\dagger}\textbf{L})(\tilde{y}) + \textbf{L}(\tilde{x})\textbf{L}^{\dagger}(\tilde{y}) \right] \\
& \quad +  \int_{0}^{t'}  dt'' \left[ M(\tilde{x}; t',t'') - M^{\dagger}(\tilde{y}; t',t'') + N(\tilde{x}; t',t'') N^{\delta}(\tilde{y}; t',t'') \right] \Big\rbrace  ,
\end{split}
\end{equation}
where L' must be the Lagrangian that comes from the operative Hamiltonian via a Legendre transformation.

The second is to rewrite the FV/CL result given by equation (25) into
\begin{equation}\label{27}
\begin{split}
J(x,y;x',y'; t)& = \int_{end points} (d\tilde{x}) (d\tilde{y}) \exp \frac{i}{\hbar} \int_{0}^{t} dt' \Big\lbrace \left[ L_{s}(\tilde{x},\dot{\tilde{x}}) - L_{s}(\tilde{y},\dot{\tilde{y}}) \right]\\
& \quad + \int_{0}^{t'} dt'' \left[ M(\tilde{x}; t',t'') - M^{\dagger}(\tilde{y}; t',t'') + N(\tilde{x}; t',t'') N^{\delta}(\tilde{y}; t',t'') \right] \Big\rbrace ,
\end{split}
\end{equation}
where
\begin{subequations}\label{28}
\begin{gather}
M(\tilde{x}; t',t'') = \tilde{x}(t') \sum_{i} \frac{c_{i}^{2}}{m\omega_{i}} \left[ i\left( \coth \frac{\hbar \omega_{i}}{kT} + \dfrac{1}{\sinh \frac{\hbar \omega_{i}}{kT}} \right) \cos \omega_{i}(t'-t'') + \frac{1}{2} \sin \omega_{i}(t'-t'') \right] \tilde{x}(t''),\label{second}\\
N(\tilde{x}; t',t'')N^{\delta}(\tilde{y}; t',t'') = \begin{pmatrix}
	                                                                                  A \tilde{x}(t') & B \tilde{x}(t'')
	                                                                                  \end{pmatrix}
	                                                                                  \cdot
	                                                                                  \begin{pmatrix}
	                                                                                  A \tilde{y}(t'')\\
	                                                                                  B \tilde{y}(t')
	                                                                                  \end{pmatrix}.
\end{gather}
\end{subequations}	
The values of A and B are read off from equation (23), which gives
\begin{subequations}\label{29}
\begin{gather}
A^{2} = \sum_{i} \frac{c_{i}^{2}}{m \omega_{i}} \left[ -i \left( \coth \frac{\hbar \omega_{i}}{kT} + \dfrac{1}{\sinh \frac{\hbar \omega_{i}}{kT}} \right) \cos \omega_{i}(t'-t'') + \frac{1}{2} \sin \omega_{i}(t'-t'') \right] ,\label{first}\\
B^{2} =  \sum_{i} \frac{c_{i}^{2}}{m \omega_{i}}\left[ -i \left( \coth \frac{\hbar \omega_{i}}{kT} + \dfrac{1}{\sinh \frac{\hbar \omega_{i}}{kT}} \right) \cos \omega_{i}(t'-t'') -  \frac{1}{2} \sin \omega_{i}(t'-t'') \right] 
\end{gather}
\end{subequations}
Comparing equations (26) with equations (27), (28) and (29) the following identification can be made 
	
	(1) The the Lagrangian L' of the Lindblad Plus is equal to the Lagrangian of the system $ L_{s} $, thus the operative Hamiltonian of the Lindblad Plus is equal to the Hmiltonian of the system.
	
	(2) Since there is no single time integral involving $ \textbf{L}(\tilde{x})\textbf{L}^{\dagger}(\tilde{y}) $ in equation (27), it follows that $ \textbf{L} = 0 $, i.e., there is no Lindbladian.
	
	(3) The non-Markov terms in the Lindblad Plus are precisely as given in equations (27) and (28).
	
	(4) The general master equation for this example is given by equation (10) with $ H_{s} $, $ \textbf{L} = 0 $, and the matrix elements of the non-Markov terms $ M(t',t'') $ and $ N(t',t'') $ and $ N^{\delta}(t',t'') $ are as given in equations (27) and (28). Also, from equation (28b), $ N^{\delta} $ is almost the transpose of N but distinctly different in that it also involves the transpose of t', t''. It is for this reason that $ N^{\delta} $ was used in equation (10).

This completes the link between Feynman-Vernon and Lindblad Plus as far as this simple example is concerned. Unfortunately, there is no Lindbladian, all there is is the Plus, the non-Markov terms. In the next section, it will be pointed out where the Lindblad part will arise.
\section{\label{sec:level5}The Stationary Phase Approximation and the Generating Functional}
There is a need for a general method to evaluate the influence functional $ \textbf{F} $ to the point that the expression for the propagator of the density matrix  can be compared with the $ J_{JP} $ of the Lindblad Plus so that the operative Hamiltonian, the Lindbladian and the non-Markov terms can be determined from the system, bath dynamics and their interaction. This is the point of this section.                              

The starting point is equation (22), evaluating the bath degrees of freedom path-integral subject to end point conditions to get the effective dynamics of the system. Following the notation at the beginning of Section III, the dynamics is further specified by the Lagrangians  
\begin{subequations}\label{30}
\begin{gather}
L_{b} = \frac{m}{2} \sum_{i = 1}^{N} \left( \dot{\tilde{X}}_{i}^{2} - V(\tilde{X}_{i}) \right),\label{first}\\
L_{int} = \sum_{i = 1}^{N} \left( \alpha_{i}(\tilde{x}_{a})\tilde{X}_{i} + \beta_{ij}(\tilde{x}_{a}) \tilde{X}_{i}\tilde{X}_{j} + ... \right) 
\end{gather}
\end{subequations}
The stationary phase approximation will consider the bath classical dynamics alone given by
\begin{equation}\label{31}
m\dfrac{d^{2}\tilde{X}_{i}}{dt'^{2}} + \dfrac{\partial V}{\partial \tilde{X}_{i}} = 0
\end{equation}
Denote the solution of equation (30) by $ \bar{X}_{i}(t'; \lambda,\kappa) $ where $ \lambda $ and $ \kappa $ are the two constants of integration of equation (30). These constants will be determined by the end point conditions $ \bar{X}_{i}(t' = 0; \lambda,\kappa) = X'_{i} $ and $ \bar{X}_{i}(t' = t; \lambda,\kappa) = X_{i} $, thus I can write the classical solution as 
\begin{equation}\label{32}
\bar{X}_{i} = \bar{X}_{i}(t'; X'_{i},X_{i}),
\end{equation}
showing explicitly the dependence on the end point conditions.

Consider the path-integral given by equation (22)
\begin{equation}\label{33}
\kappa(\tilde{x_{a}}; X_{i}', X_{i}; t) = \int_{end points} \prod _{i}(d\tilde{X}_{i}) \exp {\frac{i}{\hbar} \left[ S_{b} + S_{int} \right] }.
\end{equation}
In general, this is evaluated by expanding the $ S_{int}(\tilde{x},\tilde{X}) $. 

Defining the generating functional Z(j)
\begin{equation}\label{34}
Z(j) = \int_{end points} \prod_{i} (d\tilde{X}_{i}) \exp { \frac{i}{\hbar} \left[ S_{b}(\tilde{X}) + \int_{0}^{t} dt' j_{i}(t') \tilde{X}_{i}(t') \right] }
\end{equation}
Equation (32) can then be written as
\begin{equation}\label{35}
\kappa(\tilde{x_{a}}; X_{i}', X_{i}; t) =\left[  \exp \frac{i}{\hbar} S_{int}(\tilde{x},\frac{\hbar}{i} \dfrac{\delta }{\delta j_{i}(t')})\right] Z(j)\Big\vert_{j = 0},
\end{equation}
where $ j = 0 $ is taken at the end of all operations per term in the series expansion.
I will now use the background decomposition in evaluating $ Z(j) $. I begin with
\begin{equation}\label{36}
\tilde{X}_{i} = \bar{X}_{i} + \delta X_{i},
\end{equation}
with $ \bar{X}_{i} $ the classical solution of equation (31). Since the end points $ X'_{i} and X_{i} $ of $ \tilde{X}_{i} $ are taken cared of by $ \bar{X}_{i} $, the fluctuations $ \delta X_{i} $ follow the end point conditions $ \delta X_{i}(t' = 0) = \delta X_{i}(t' = t) = 0 $. This is important in deriving the relevant Green's function later.
Using equation (36) in equation (34) results in
\begin{equation}\label{37}
\begin{split}
Z(j)& \propto \exp {\frac{i}{\hbar} \sum_{i} \left\lbrace  \frac{m}{2}  \left[ X_{i} \dot{\bar{X_{i}}}(t) - X'_{i} \dot{\bar{X_{i}}}(0) \right] + \int_{0}^{t} dt'  \bar{X}_{i}(t') j_{i}(t') \right\rbrace } \\  &\quad \int_{end points} \prod_{i} (d\delta X_{i}) \exp { \frac{i}{\hbar} \sum_{i,j}  \int_{0}^{t} dt' (-\frac{1}{2}) \left[ \delta X_{i} \left( m \dfrac{d^{2}}{dt'^{2}} \delta_{ij} + \dfrac{ \partial^{2} V}{\partial \tilde{X}_{i} \partial \tilde{X}_{j}}\Big\vert_{\tilde{X} = \bar{X}} \right) \delta X_{j}(t') +  j_{i}(t') \delta X_{i} \right] }.
\end{split}
\end{equation}
It must be emphasized that equation (32) must be used in the above equation so $ Z(j) $ is found to be end points dependent. 
Doing the $ (\delta X_{i}) $  path-integral, the generating functional becomes
\begin{subequations}\label{38}
\begin{gather}
Z(j) \propto det^{-\frac{1}{2}}G \exp {\frac{i}{\hbar}  \left\lbrace  \tilde{Z}+ \tilde{L}+ \tilde{Q} \right\rbrace } ,\label{first}\\
\tilde{Z} =  \frac{m}{2} \sum_{i} \left( X_{i} \dot{\bar{X_{i}}}(t) - X'_{i} \dot{\bar{X_{i}}}(0) \right),\label{second}\\
\tilde{L} = \sum_{i }\int_{0}^{t} dt'  \bar{X}_{i}(t') j_{i}(t'),\label{third}\\
\tilde{Q} = \int_{0}^{t} dt' \int_{0}^{t'} dt'' j_{i}(t') G_{ij}(t',t'') j_{j}(t''),
\end{gather}
\end{subequations}
where the Greens function is solved from 
\begin{equation}\label{39}
\left( m \dfrac{d^{2} }{dt'^{2}} \delta_{ij} + \dfrac{ \partial^{2} V}{\partial \tilde{X}_{i}(t') \partial \tilde{X}_{j}(t')}\Big\vert_{\tilde{X} = \bar{X}} \right) G_{jk}(t',t'') = \delta_{ik} \delta(t'-t'').
\end{equation}
The notation used shows the source dependence of the terms, i.e., none for $ \tilde{Z} $, linear for $ \tilde{L} $ and quadratic for $ \tilde{Q} $.

Equation (38) is then substituted in equation (35) to give $ \kappa(\tilde{x_{a}}; X_{i}', X_{i}; t) $. Then the same calculations given in equations (35) to (39) is done for $ \kappa^{*}(\tilde{y_{a}}; Y_{i}', X_{i}; t) $. To compute for the influence functional given by equation (20), the initial bath density function  $ \rho_{b}(x',y'; 0) $ must be known. It is in the calculation of the influence functional $ \textbf{F} $ where the Lindbladian will arise. It comes from a single time integral $ \int_{0}^{t} ... $ in the exponential that arises out of the end point integrations that will give the Lindbladian terms and the additional term(s) that will modify the system's Hamiltonian to give the operative Hamiltonian. The double time integrals $ \int_{0}^{t} dt'\int_{0}^{t'} ... $ will give the non-Markov terms. When the influence functional is substituted in equation (19) to give the propagator for the density function J, the result, after mathematical rearrangement of terms should then be compared to the Lindblad Plus path-integral given by equation (13), and should give expressions for the operative Hamiltonian, the Lindbladian $ \textit{L} $ and the non-Markov terms M and N. 

To show that this procedure works, I will apply it to the CL/FV example discussed in the previous section. In this case, the background decomposition given in equation (36) involves the harmonic equation
\begin{equation}\label{40}
\dfrac{d^{2}\bar{X}_{i}}{dt'^{2}} + \omega_{i}^{2} \bar{X}_{i} = 0,
\end{equation} 
with solution 
\begin{subequations}\label{41}
\begin{gather}
\bar{X}_{i} = A_{i} \sin \omega_{i}t' + B_{i} \cos \omega_{i}t', \label{first}\\
B_{i} = X'_{i}, \label{second}\\
A_{i} = \dfrac{1}{\sin \omega_{i}t} X_{i} - \cot \omega_{i}t X'_{i}.
\end{gather}
\end{subequations}
The constants $ A_{i} $ and $ B_{i} $ are determined from the end point conditions $ \tilde{X}_{i}(t'=0) = \bar{X}_{i}(t'=0) = X'_{i} $,  $ \tilde{X}_{i}(t'=t) = \bar{X}_{i}(t'=t) = X_{i} $. This gives the end point conditions for $ \delta X_{i}(t'=0) = \delta X_{i}(t'=t) = 0 $, which must be satisfied by the Greens function of $ \dfrac{d^{2}}{dt'^{2}} + \omega_{i}^{2}  $. Taking all these into account, the generating functional is
\begin{subequations}\label{42}
\begin{gather}
Z(j)  \propto \exp { \frac{i}{\hbar} \left[ \tilde{Z} + \tilde{L} + \tilde{Q} \right] },\label{first}\\
\tilde{Z} = \sum_{i} \left[ \frac{m\omega_{i}}{2} \cot \omega_{i}t \left( X_{i}^{2} + X_{i}^{\prime 2} \right)  - \dfrac{m\omega_{i}}{\sin \omega_{i}t} X_{i}X'_{i} \right]  , \label{second}\\
\tilde{L} = \sum_{i} c_{i} \int_{0}^{t} dt' \left[ \dfrac{X_{i}}{\sin \omega_{i}t } \sin \omega_{i}t' j_{i}(t') + X'_{i} \left(  \cos \omega_{i}t' - \cot \omega_{i}t  \sin \omega_{i}t' \right)  j_{i}(t') \right] ,\label{third}\\
\tilde{Q}= \sum_{i} c_{i}^{2} \int_{0}^{t} dt' \int_{0}^{t'} dt'' j_{i}(t')\left[  \dfrac{1}{m\omega_{i} \sin \omega_{i}t } \sin \omega_{i}(t - t') \sin \omega_{i}t'' \right] j_{i}(t''),
\end{gather}
\end{subequations}
where the determinant factor in equation (38a) was neglected because it is not end points $ (X_{i}, X'_{i}) $ dependent for this example. That this gives the same result as the previous section only needs to show that equations (35) and (42) reproduce equation (23) exactly. And indeed it does. Once equation (23) is reproduced, the discussions in the previous section from equation (24) to equation (29) and the Appendix shows how that the Feynman-Vernon theory is equivalent to Lindblad Plus, which for this example unfortunately gives the Plus only as there are no single time-integrals in the exponential that arise from the $ \textbf{F} $.   

This section shows that the stationary phase approximation method in the presence of a source j, can be used to compute for $ \kappa(\tilde{x}; X', X; t) $, from which the influence functional $ \textbf{F} $ and the propagator for the density matrix J can be computed.  The hard part is in rearranging the terms to make the identification of the result to the path-integral of Lindblad Plus.  

\section{\label{sec:level6}Conclusion}
This work proposed how to generalize the LGKS formalism for open systems to include memory effects in equation (10). The path-integral equivalent of the integro-differential general master equation is then derived resulting in the propagator for the density matrix $ J_{LP} $ given by equation (13). 

At the other end, generally, open systems, are described by giving the system's dynamics, the dynamics of the bath/environment and specifying how the system and bath interacts. This is the purview of the Feynman-Vernon method, which integrates out the bath degrees of freedom to give the propagator for the density matrix $ J_{FV} $ as given by equation (19). In this work, I introduced the stationary phase method in the presence of a source j to define the generating functional $ Z(j) $ and use it in computing the $ J_{FV} $ in a way where the single time integrals (operative Hamiltonian and Lindbladian) and double time integrals (memory effects) are transparent. Thus when $ J_{FV} $ is compared to $ J_{LP} $, the terms that appear in equation (10) are determined from the system, bath and interaction dynamics. 

Applying the formalism to the simple example discussed by FV and CL, these terms are indeed determined. Unfortunately, the operative Hamiltonian of the system is just the pure system Hamiltonian, the Lindbladian is zero and the memory terms are given in equations (28) and (29). 

A natural extension of this paper is to apply the formalism here in another system/bath interaction that will yield a non-trivial Lindbladian and the system's operative Hamiltonian changed by the bath degrees of freedom aside from giving the memory effect terms (M and N). 
  
\begin{acknowledgements}
Mike Solis of the NIP helped me in sourcing some of the references and showed interest in the early part of the work. I am grateful to him for his help and discussions. I appreciate the help of Ms. Antonieta Villaflor of the College of Science Library of the University of the Philippines for providing me with some reference materials. I would like to thank Felicia Magpantay for correcting my Latex file and Gravity for patiently keeping me company while I worked on this paper. 
\end{acknowledgements}

\begin{appendices}

\renewcommand{\theequation}{A-\arabic{equation}}
\setcounter{equation}{0}  
\setcounter{section}{0}
\section*{Appendix. Details of Computation for Equation (25)}
\setcounter{section}{0}
This appendix is necessitated by the fact that the answer here differs slightly from the published literature \cite{Caldeira}. First, I note that that the quoted FV result in equation (23) differs from the cited equation (4.7b) because of the difference in the sign of the interaction term given in equation (21b) compared to FV. The sign used here follows CL. 

Using equation (23) and a similar term for $ \kappa^{*}(\tilde{y}; Y', X; t) $, the influence functional given by equation(20) 
\begin{equation}\label{A.1}
\textbf{F}(\tilde{x},\tilde{y}) = \int dX dX' dY'  \kappa(\tilde{x}; X', X; t) \rho_{b}(X',Y'; 0) \kappa^{*}(\tilde{y}; Y', X; t), 
\end{equation}
where $ \rho_{b} (X',Y'; 0) $ is given by equation (24). The terms in the exponential in above, aside from the overall factor $ \frac{i}{\hbar} $, are either quadratic, linear or zeroth order in the end point conditions. Thus the integrations are doable. These are
\begin{subequations}\label{A.2}
\begin{gather}
zeroth =  \frac{1}{m \omega_{i}} \dfrac{1}{\sin \omega_{i}t} c_{i}^{2} \int_{0}^{t} dt' \int_{0}^{t'} dt''  \sin \omega_{i}(t - t') \sin \omega_{i}t'' [ \tilde{x}(t') \tilde{x}(t'') - \tilde{y}(t') \tilde{y}(t'') ],\label{first}\\
linear =  \sum_{i} c_{i} \left\lbrace  \frac{X_{i}}{\sin \omega_{i}t}  \int dt' [ \tilde{y}(t') - \tilde{x}(t') ] \sin \omega_{i}t' + \int dt' \left( \cos \omega_{i}t' - \cot \omega_{i}t \sin \omega_{i}t' \right) [Y'_{i} \tilde{y}(t') - X'_{i} \tilde{x}(t') ] \right\rbrace ,\label{second}\\
\begin{split}
quadratic& = \sum_{i} \Big\lbrace \frac{m\omega_{i}}{2} \cot \omega_{i}t \left[ (X_{i}^{2} + X_{i}^{\prime 2} ) - ( Y_{i}^{\prime 2} + X_{i}^{2} ) \right] - \frac{m\omega_{i}}{\sin \omega_{i}t} \left[ X_{i}X'_{i} - X_{i}Y'_{i} \right] \\
& \quad + i \dfrac{m\omega_{i}}{2 \sinh \frac{\hbar \omega_{i}}{kT}} \left[ ( X_{i}^{\prime 2} + Y_{i}^{\prime 2} ) \cosh \frac{\hbar \omega_{i}}{kT} - 2 X'_{i}Y'_{i} \right] \Big\rbrace.
\end{split}
\end{gather}
\end{subequations}
Since the $ X_{i}^{2} $ terms cancel out, the $ X_{i} $ integration yields a delta function giving
\begin{equation}\label{A.3}
Y'_{i} = X'_{i} + \frac{c_{i}}{m\omega_{i}} \int_{0}^{t} dt' \left[ \tilde{x}(t') - \tilde{y}(t') \right] \sin \omega_{i}t'.
\end{equation}
Substituting equation (A.3) in equations (A.1) and (A.2), the resulting $ X'_{i} $ integration is of the following form
\begin{subequations}\label{A.4}
\begin{gather}
X'_{i } integral = \int dX'_{i} \exp {\frac{i}{\hbar} \left( \alpha X_{i}^{\prime 2} + \beta X'_{i} + \gamma \right)} ,\label{first}\\
\frac{i}{\hbar} \alpha = -\dfrac{m\omega_{i}}{\hbar \sinh \frac{\hbar \omega_{i}}{kT}} \left[ \cosh \frac{\hbar \omega_{i}}{kT} - 1 \right] ,\label{second}\\
\beta = c_{i} \int_{0}^{t} dt' \left[ \tilde{y}(t') - \tilde{x}(t') \right]  \cos \omega_{i}t' + i \dfrac{c_{i}}{\sinh \frac{\hbar \omega_{i}}{kT}} \left[ \cosh \frac{\hbar \omega_{i}}{kT} - 1 \right] \eta , \label{third}\\
\eta = \int_{0}^{t} dt' \left[ \tilde{x}(t') - \tilde{y}(t') \right] \sin \omega_{i}t',
\end{gather}
\end{subequations}
and $ \gamma $ is given by
\begin{equation}\label{A.5}
\begin{split}
\gamma &=\Big\lbrace  -\frac{c_{i}^{2}}{2m\omega_{i}} \eta^{2} \cot \omega_{i}t  + \frac{c_{i}^{2}}{m\omega_{i}} \eta \int_{0}^{t} dt' \tilde{y}(t') \left[ \cos \omega_{i}t' - \cot \omega_{i}t \sin \omega_{i}t' \right] \\
& \quad + i \frac{c_{i}^{2}}{m\omega_{i}} \eta^{2} \coth \frac{\hbar \omega_{i}}{kT}  \Big\rbrace .
\end{split}
\end{equation}
From equation (A.4b), this integration is well-defined. The result is 
\begin{equation}\label{A.6}
end points integrals = \exp {\frac{i}{\hbar} \left[ \gamma - \frac{1}{4} \dfrac{\beta^{2}}{\alpha} \right]},
\end{equation}
where we have neglected the determinant factor since it is not in the exponential. Still to be added to equation (A.6) is the zeroth term in equation (A.2). Observe that equation (A.6) involves double integrations of the form $ \int_{0}^{t} dt' \int_{0}^{t} dt'' f(t',t'') $ while the zeroth term of equation (A.2) involves integrations of the form $ \int_{0}^{t} dt' \int_{0}^{t'} dt'' f(t',t'') $. Converting the (A.6) integrals to the form of the zeroth term of (A.2) makes use of the following integration rule given in the Appendix of Feynman and Hibbs \cite{FeynmanR}.
\begin{equation}\label{A.7}
\int_{0}^{t} dt' \int_{0}^{t'} dt'' f(t',t'') = \int_{0}^{t} dt' \int_{t'}^{t} dt'' f(t'',t') 
\end{equation}
There is one more relationship needed to complete the derivation of the influence functional that will give the J of equation (25) and this is
\begin{equation}\label{A.8}
\left[ \coth \frac{\hbar \omega_{i}}{kT} - \dfrac{1}{\sinh \frac{\hbar \omega_{i}}{kT}} \right] ^{-1} = \left[ \coth \frac{\hbar \omega_{i}}{kT} + \dfrac{1}{\sinh \frac{\hbar \omega_{i}}{kT}} \right]. 
\end{equation}
That this is true is shown by the fact that it leads to the identity $ \cosh ^{2} \frac{\hbar \omega_{i}}{kT} - \sinh ^{2} \frac{\hbar \omega_{i}}{kT} = 1 $.
All these give 
\begin{equation}\label{A.9}
\begin{split}
\textbf{F}(\tilde{x},\tilde{y})& = \exp  \frac{i}{\hbar} \Bigg\lbrace \int_{0}^{t} dt' \int_{0}^{t'} dt''  [ \tilde{x}(t') - \tilde{y}(t') ] \Big( i \sum_{i} \dfrac{c_{i}^{2}}{m\omega_{i}} \Big( \coth \frac{\hbar \omega_{i}}{kT}\\
& \quad + \dfrac{1}{\sinh \frac{\hbar \omega_{i}}{kT} } \Big) \cos \omega_{i}(t' - t'') \Big) [ \tilde{x}(t'') - \tilde{y}(t'') ] + \int_{0}^{t} dt' \int_{0}^{t'} dt'' [ \tilde{x}(t') - \tilde{y}(t') ] \Big( \sum_{i} \frac{ c_{i}^{2}}{2m\omega_{i}} \sin \omega_{i}(t' - t'') \Big) [ \tilde{x}(t'') + \tilde{y}(t'') ] \Bigg\rbrace ,
\end{split}
\end{equation}
which gives the propagator of the density matrix given by equation (25).

This completes the proof made in Section IV.
 

\end{appendices}
\end{document}